\documentclass[doublespacing]{elsart}
\usepackage{graphicx}
\usepackage{bm} 
\usepackage{amssymb}
\usepackage{lineno}
\journal{Nucl. Instr. and Meth. in Phys. Res. A}
\begin{document}
\begin{frontmatter}
\title {A large acceptance scintillator detector
with wavelength shifting fibre read-out for search of eta-nucleus bound states}
\author{The GEM Collaboration:}
\author[a]{M.G.Betigeri},
\author[b]{P.K.Biswas},
\author[c]{A.~Budzanowski},
\author[a]{A.~Chatterjee},
\author[d]{R.~Jahn},
\author[b]{S.Guha},
\author[e]{P.Hawranek},
\author[n]{B.K.Jain},
\author[b]{S.B.Jawale},
\author[a]{V.~Jha\corauthref{cor1}}
\ead{vjha@barc.gov.in}
\corauth[cor1]{Corresponding author. Nuclear Physics Division, BARC, Trombay,
Mumbai-400085, India. Tel:++91-22-25593457;
Fax: ++91-22-25505151.},
\author[f]{K.~Kilian},
\author[c]{S.~Kliczewski},
\author[f]{Da.~Kirillov},
\author[g]{Di.~Kirillov},
\author[h]{D.~Kolev},
\author[i]{M.~Kravcikova},
\author[j]{T.~Kutsarova},
\author[f]{M.~Lesiak},
\author[k]{J.~Lieb},
\author[f,o]{H.~Machner},
\author[e]{A.~Magiera},
\author[f]{R.~Maier},
\author[l]{G.~Martinska},
\author[m]{S.~Nedev},
\author[g]{N.~Piskunov},
\author[f]{D.~Prasuhn},
\author[f]{D.~Proti\'c},
\author[f]{J.~Ritman},
\author[f]{P.~von Rossen},
\author[a]{B.~J.~Roy},
\author[a]{P.~Shukla},
\author[g]{I.~Sitnik},
\author[c,f]{R.~Siudak},
\author[h]{R.~Tsenov},
\author[l]{M.~Ulicny},
\author[l]{J.~Urban},
\author[f,h]{G.~Vankova}

\address[a]{Nuclear Physics Division, BARC, Mumbai-400 085, India}
\address[b]{Centre for Design and Manufacture, BARC, Mumbai-400 085, India}
\address[c]{Institute of Nuclear Physics, Polish Academy of Sciences, Krakow, Poland}
\address[d]{Helmholtz-Institut f\"{u}r Strahlen- und Kernphysik der Universit\"{a}t Bonn, Bonn, Germany}
\address[e]{Institute of Physics, Jagellonian University, Krakow, Poland}
\address[f]{Institut f\"{u}r Kernphysik, Forschungszentrum J\"{u}lich, J\"{u}lich, Germany}
\address[g]{Laboratory for High Energies, JINR Dubna, Russia}
\address[h]{Physics Faculty, University of Sofia, Sofia, Bulgaria}
\address[i]{Technical University Kosice, Kosice, Slovakia}
\address[j]{Institute of Nuclear Physics and Nuclear Energy, Sofia, Bulgaria}
\address[k]{Physics Department, George Mason University, Fairfax, Virginia, USA}
\address[l]{P.~J.~Safarik University, Kosice, Slovakia}
\address[m]{University of Chemical Technology and Metallurgy, Sofia, Bulgaria}
\address[n]{Physics Department, Mumbai University, Vidyanagari, Mumbai}
\address[o]{Fachbereich Physik, Universit\"{a}t Duisburg-Essen, Duisburg, Germany}
\begin{abstract}

   A large acceptance scintillator detector with
wavelength shifting optical fibre readout has been designed and
built to detect the decay particles of $\eta$-nucleus bound system
(the so-called $\eta$-mesic nuclei), namely, protons and pions. The
detector, named as ENSTAR detector, consists of 122 pieces of
plastic scintillator of various shapes and sizes, which are arranged
in a cylindrical geometry to provide particle identification, energy
loss and coarse position information for these particles. A solid
angle coverage of $\sim$95\% of total 4$\pi$ is obtained in the
present design of the detector. Monte Carlo phase space calculations
performed to simulate the formation and decay of $\eta$-mesic nuclei
suggest that its decay particles, the protons and pions are emitted
with an opening angle of 150$^\circ \pm 20^\circ$, and with energies
in the range of 25 to 300 MeV and 225 to 450 MeV respectively. The
detailed GEANT simulations show that $\sim$ 80 \% of the decay
particles (protons and pions) can be detected within ENSTAR. Several
test measurements using alpha source, cosmic-ray muons etc. have
been carried out to study the response of ENSTAR scintillator
pieces. The in-beam tests of fully assembled detector with proton
beam of momentum 870 MeV/c from the Cooler synchrotron COSY have
been performed. The test results show that the scintillator fiber
design chosen for the detector has performed satisfactorily well.
The present article describes the detector design, simulation
studies, construction details and test results.
\end{abstract}
\end{frontmatter}
\begin{linenumbers}
\section{Introduction}
             A large acceptance plastic scintillator detector ENSTAR has been designed and built
for studies of
$\eta$-mesic nuclei -
         a bound system of  $\eta$-meson and a nucleus.
         The finding of strong and attractive nature of the $\eta$-nucleon($\eta$-N) scattering length
and the presence of a resonance near the $\eta$-N threshold,
provide an interesting possibility of the formation of $\eta$-nucleus bound states
          \cite{LIU,BHALERAO}. The experimental confirmation of the existence of such bound
         systems would open up new avenues for elucidation of the $\eta$-nucleus
dynamics at intermediate energies.
Such experiments \cite{BEAM} are being performed at the intermediate energy accelerator
 facility COSY J\"{u}lich, using GeV energy proton beam. The experiments use recoil-free transfer reactions
         p+($^Z$X${_A}$) $\rightarrow$  $^3$He + $(^{Z-1}X_{A-2})_\eta$ on several target nuclei X = Li, C, Al, etc.
         The expected cross section for events corresponding to formation of $\eta$-mesic nuclei
         is rather low, hence, a dedicated detection system is needed to enhance the sensitivity of
the measurement.  ENSTAR is the part of detection
         system which has been developed in order to obtain an unambiguous signal for the formation and decay of the
         $\eta$-nucleus bound state. The outgoing $^3$He particles are detected in the Big Karl detection system \cite{Martin,Gemnim},
          which includes a magnetic spectrograph and its focal plane detectors consisting of drift chambers and
         scintillator hodoscopes. The corresponding proton and pion from the decay of $\eta$-mesic nucleus are
         registered in ENSTAR.
          In addition to the  $\eta$-bound states search, the ENSTAR detector
          can also be used in many other experiments where the missing mass determination
          of the reaction product needs to be done in coincidence with its decay products e.g., for the study of
          $\Delta$ interaction in nuclear matter, where the decay products
          of $\Delta$ states, protons and pions can be detected by ENSTAR \cite{albert1}.
 The details of the Big Karl
spectrometer have been reported elsewhere \cite{Martin,Gemnim}. In
this paper, the description of the newly built ENSTAR detector is
reported. The geometric design, simulation studies and fabrication
procedure are described. Test measurements done at various stages
during the construction of ENSTAR as well as the in-beam tests
performed at the COSY accelerator are presented.
\section{Physics background and ENSTAR design considerations}
         Phase space calculations  to simulate eta-mesic nucleus decay events
         were performed using the N-body Monte-Carlo
         event generator program ``Genbod'' \cite{GENBOD}. The program generates multi-particle
          weighted events according to Lorentz invariant Fermi phase space.
         The reaction p+$^{16}$O $\rightarrow$ $^3$He+$^{14}N_{\eta}$ was studied at a momentum close
         to the magic momentum. The magic momentum is defined as the beam momentum at which recoil-less $\eta$
         can be produced in the elementary process. For the reaction considered, the elementary reaction is
          pd $\rightarrow$ $^3$He$\eta$, for which the magic momentum was calculated to be 1.745 GeV/c,
        corresponding to a proton kinetic energy of T$_p$=1.05 GeV.
         The $\eta$-nuclei formation proceeds through the excitation of N$^*$ (1535 MeV) resonance and one of its decay
        channels is through proton and pion. The simulations were performed in two steps.
        In the first step, Monte Carlo events were generated for the
        p+$^{16}$O $\rightarrow$ $^3$He+$^{14}$N$_{ex}$ reaction where an excitation energy of 547 MeV,
        equal to the mass of eta meson, is given to $^{14}$N nucleus. Only those
        $^{14}$N events were considered  for which the corresponding $^3$He particle is within the Big Karl
acceptance ($\theta_{lab}(^3$He$) \leq 6^\circ$). In the next step, the decay of N$^*$ to
p-$\pi$ pair was simulated. The mass of N$^*$ was taken equal to the mass of a nucleon plus the
mass of an eta meson, while its velocity was assumed to be the same as that of
the recoil  $^{14}$N modified  by the Fermi momentum distribution.
The p-$\pi$ opening angle distribution
shows a  peak  at around $\approx$150$^\circ$ with a width of 40$^\circ$ (Fig. 1). The
energy spectrum for the proton peaks at T$_p$ $\approx$100 MeV with a width
(FWHM) of 120 MeV (Fig. 2),
 while the pion spectrum has a peak at $\approx$320 MeV and a
similar width(Fig. 3) as that of proton peak. The simulations were
also carried out for other eta-mesic nuclei formation reactions on
different target nuclei. The energy spectra and opening angle
distributions were found to be similar as that in the previous
case.\\

\begin{figure}[!ht]
\begin{center}
\includegraphics[scale=0.5]{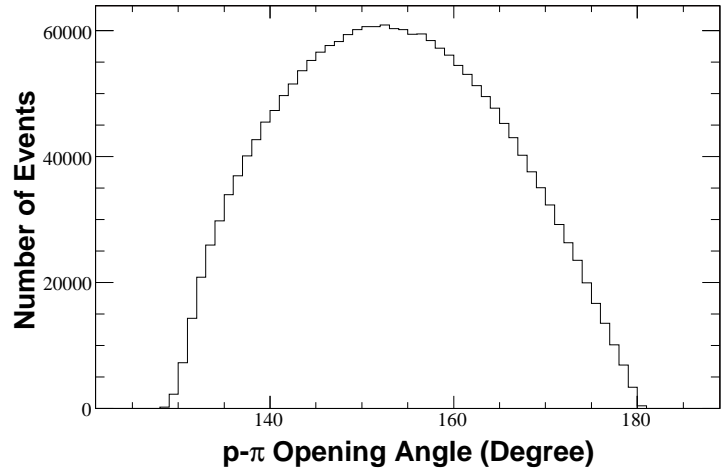}
\caption{The p-$\pi$ opening angle distribution for $\eta$-mesic
nucleus decay particles obtained from Monte-Carlo phase space
calculations as detailed in the text.}
\end{center}
\end{figure}

\begin{figure}[!ht]
\begin{center}
\includegraphics[scale=0.5]{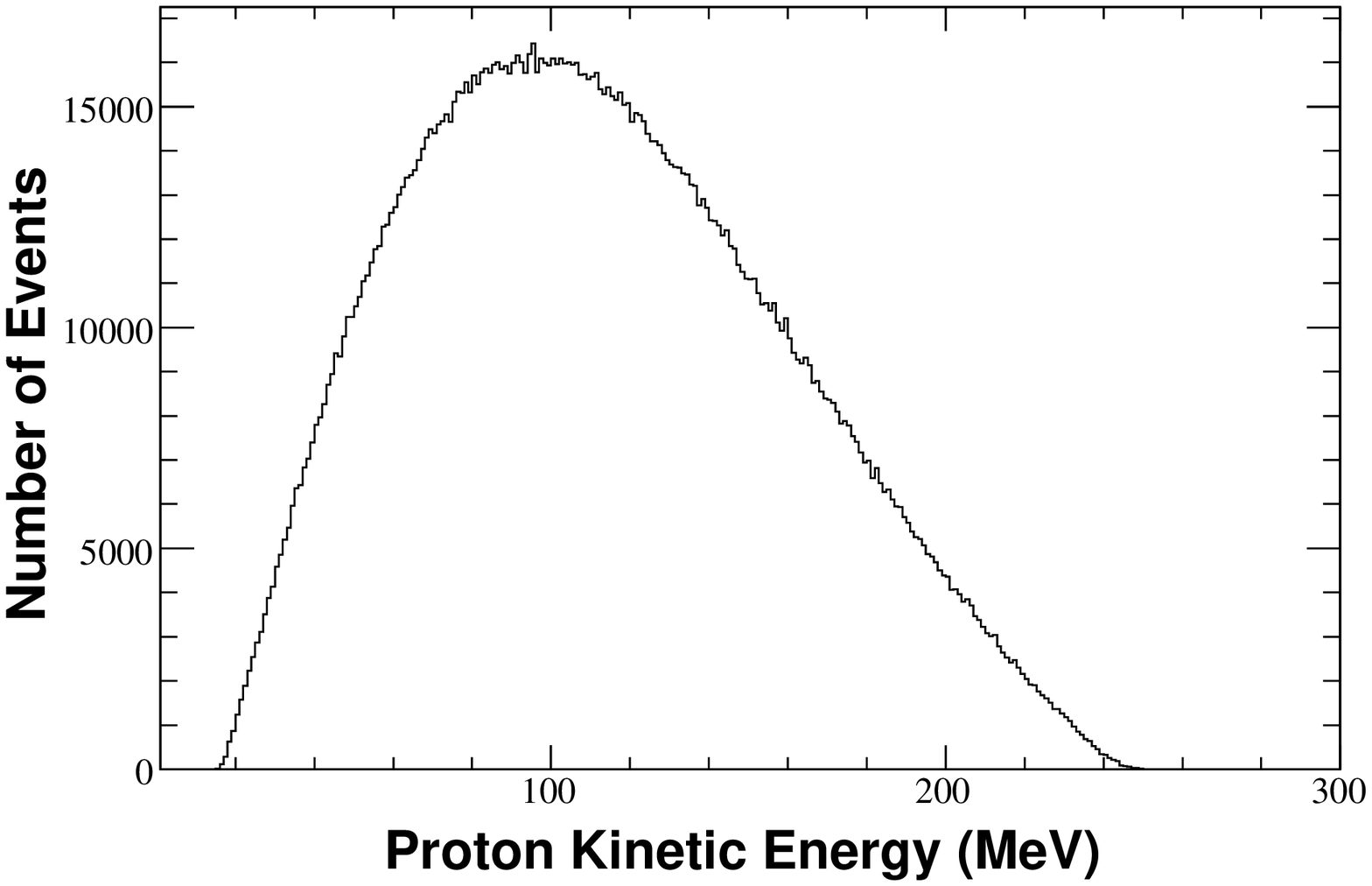}
\caption{Kinetic energy distribution of protons from $\eta$-mesic
nucleus decay obtained from Monte-Carlo phase space calculations.}
\end{center}
\end{figure}

\begin{figure}[!ht]
\begin{center}
\includegraphics[scale=0.5]{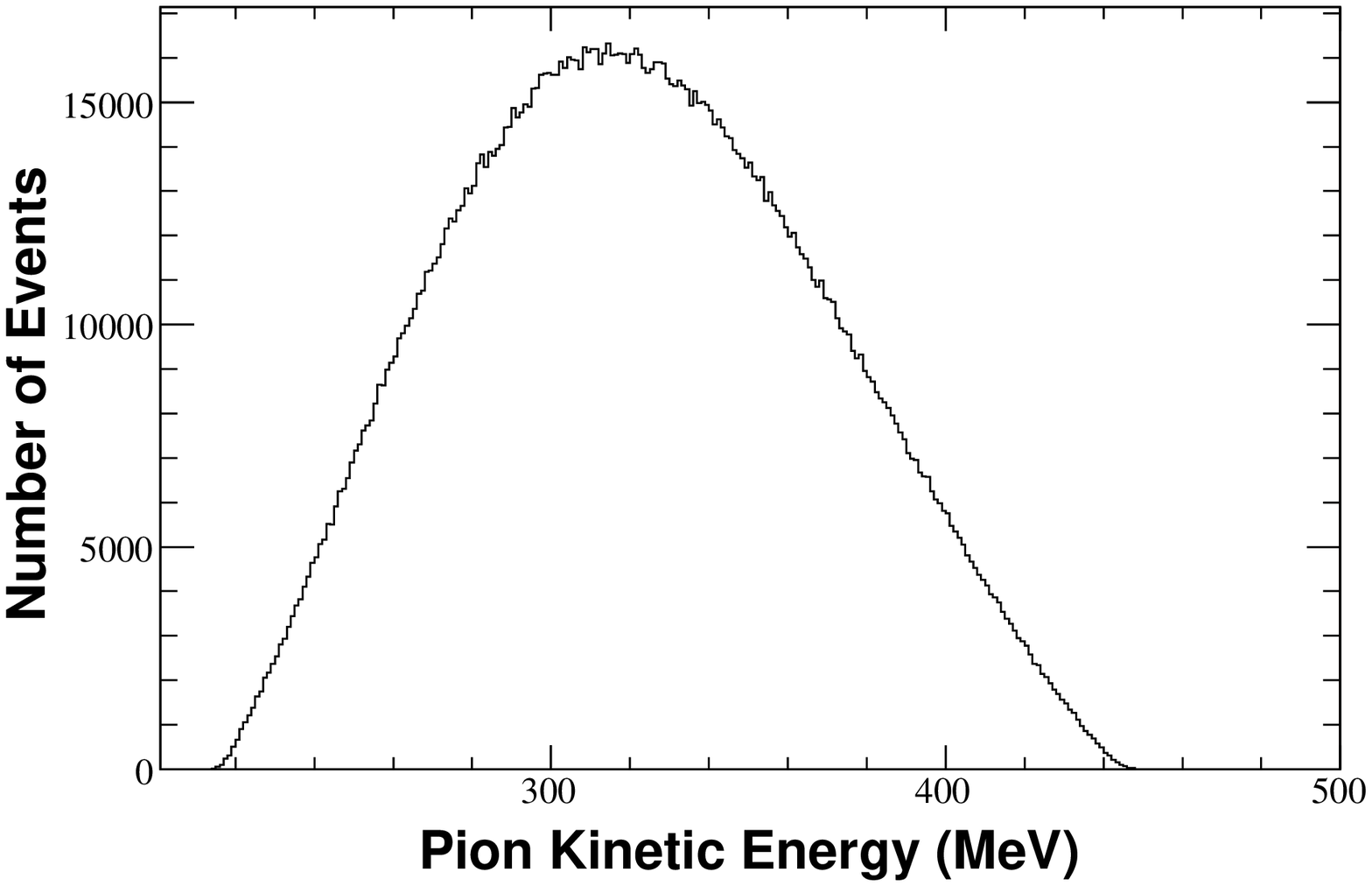}
\caption{Kinetic energy distribution of pions from $\eta$-mesic
nucleus decay obtained from Monte-Carlo phase space calculations.}
\end{center}
\end{figure}

A detector employing plastic scintillators in the $\Delta$E - E
configuration,  which provides the particle identification and
energy information of the measured particles, has been chosen for
the present design. The thickness of the detector elements has been
designed to stop the decay protons and obtain a good signal for
pions, keeping in mind the space constraints around the detector in
the experimental area. The detector has been segmented in both
$\theta$ and $\phi$
 direction for obtaining position
information with the desired granularity. Large solid angle coverage
has been achieved by minimising any unwanted material within the
detector.

\section {Design details and fabrication}
\subsection {Detector geometry}
\begin{figure}[!t]
\begin{center}
\includegraphics[scale=0.5]{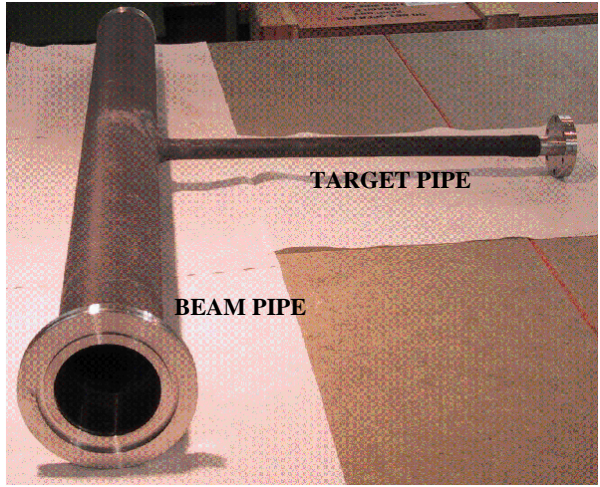}
\caption{Photograph of the scattering chamber made from the carbon
fibre material. It consists of a beam pipe and a thin target pipe
for inserting a target ladder. } \label{scatt}
\end{center}
\end{figure}

\begin{figure}[!ht]
\parbox[c]{0.50\textwidth}{
\centering
\includegraphics[width=0.5\textwidth]{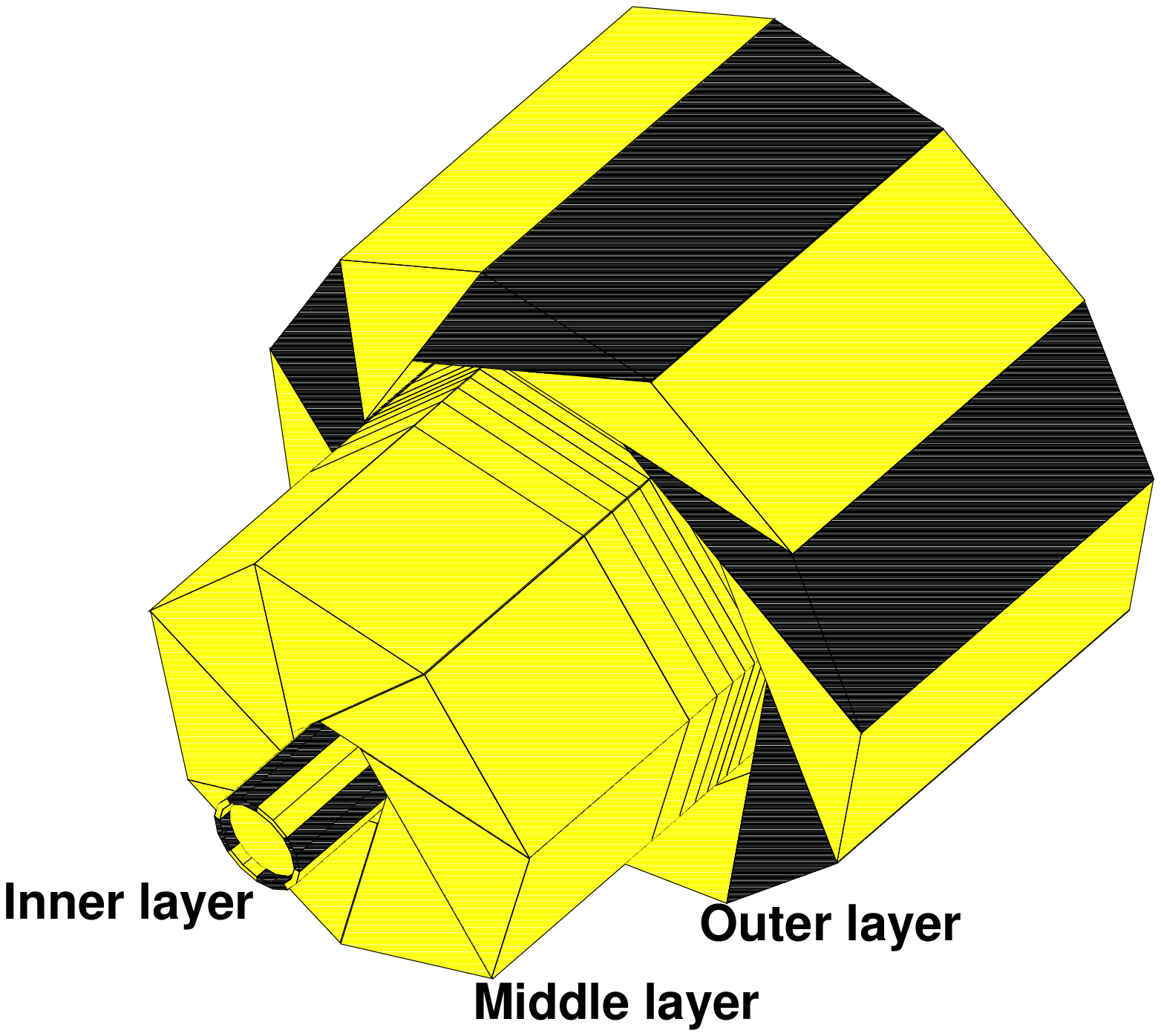}
}
\parbox[c]{0.42\textwidth}{
\centering
\includegraphics[width=0.45\textwidth]{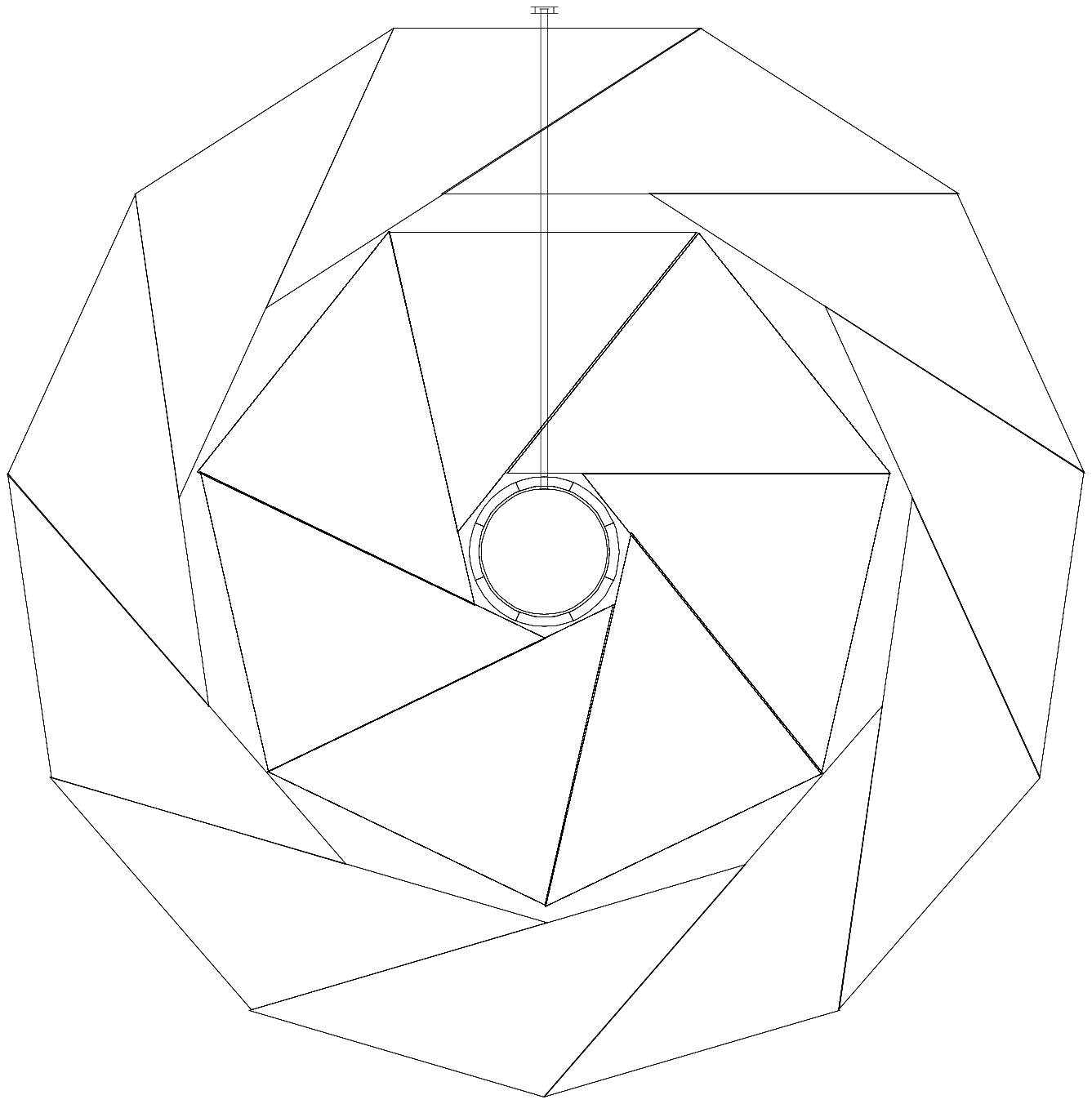}
} \caption{Left part : An assembly drawing of ENSTAR detector is
shown. Some pieces of the middle and outer layers are moved out for
an inside view. Detailed dimensions are given in the text. Right
part : A Sectional view of the detector through the target is shown.
Beam pipe along with the target pipe attached to it, is also drawn.}
\label{geom}
\end{figure}

  Based on the design and geometric criteria, ENSTAR is cylindrically shaped
with three layers of plastic
scintillators. These layers are used to generate $\Delta E- E$
spectrum for particle identification and to obtain total energy
information for the stopped particles. Each layer is divided
into a number of pieces to obtain $\theta$ and $\phi$ information.
The detector, which is made up of two identical half cylinders, is assembled
 around a scattering chamber of 1.5 mm thick carbon compound fibre
material. The scattering chamber as shown in Fig.~\ref{scatt} is
designed in a "T" shape with a thin target pipe projecting out from
the middle of beam pipe. The two half cylinders of the detector are
placed on either side of the target pipe. The target pipe has
sufficient space from inside to enable mounting of solid targets. A
Liquid target chamber, similar to the one existing at COSY
laboratory can also be used after some modifications. The angular
coverage of  the detector is
  $\theta_{lab}$ = 15$^o-165^o$ in the $\theta$-direction, while its cylindrical geometry ensures an azimuthal
 angle coverage of  $\phi=0^o-360^o$. With the
present design, the detector provides a solid angle coverage of
$\sim$95\% of 4$\pi$.
 An assembly  drawing  of ENSTAR  together with its sectional view through
the target is shown in Fig.~\ref{geom}. A total of 122 pieces of
scintillators of different shapes and dimensions are used to give
three concentric cylindrical layers on assembly.\\

The inner layer is used to provide the energy loss and  $\phi$
information of the particles passing through  it  and  is  designed
as two hollow plastic scintillator cylinders  with the following
dimensions; Inner diameter(ID) = 84 mm, Outer diameter(OD)=96 mm and
a length of 390 mm. Both the cylinders  are  split into eight  equal
sectors with a sector angle equal to 45$^o$.  Thus the  inner  layer
consists  of  a  total  of 16 segmented  annuli each of which is
read out separately.   A ~$\phi$ resolution of 45$^o$ is
satisfactory for  the  studies on  $\eta$ -mesic nuclei, as the
decay particles are emitted with a very large opening angle between
them. Signals  from  the middle layer are used to obtain energy and
$\theta$ information. This layer  consists of seven identical
scintillator bars in both the halves, each in the form of an
isosceles triangle with base = 243.1 mm
 and height = 152.4 mm
arranged to form an annular cylinder of ID=100 mm, OD= 449.4 mm and
length = 390 mm in each half. Each of triangular bars (390 mm long)
is further split lengthwise into six pieces of length 13 mm, 16 mm,
21 mm, 37 mm, 213 mm, and 90 mm so that  each piece covers an angle
interval of $ \Delta \theta_{lab}$  equal to 15$^o$. A total of 84
pieces of scintillators are used for the middle layer cylinder. The
geometrical granularity allows an angular resolution  of $\Delta
\theta_{lab}$  equal to 15$^o$.
   In conjunction  with signals from middle layer, signals from the outer layer are expected
    to provide an unambiguous signal for pions. The outer layer consists of a total  of
 22 identical bars, each 390 mm long and
a cross section of an isosceles triangle with base = 328.3 mm and
height = 105.5 mm. These outer layer pieces form an annular cylinder
of  ~~ID = 453.5   mm,  ~OD = 692.5  mm. Thus, with two identical
cylinders on either side of the target for all the three layers, the
detector  provides an angular coverage of  $15 \leq \theta_{lab}
\leq 165^o$ in the $\theta$-direction and almost full coverage in
the $\phi$-direction.\\

\subsection{GEANT simulation }
\begin{figure}[!ht]
\begin{center}
\includegraphics[scale=0.6]{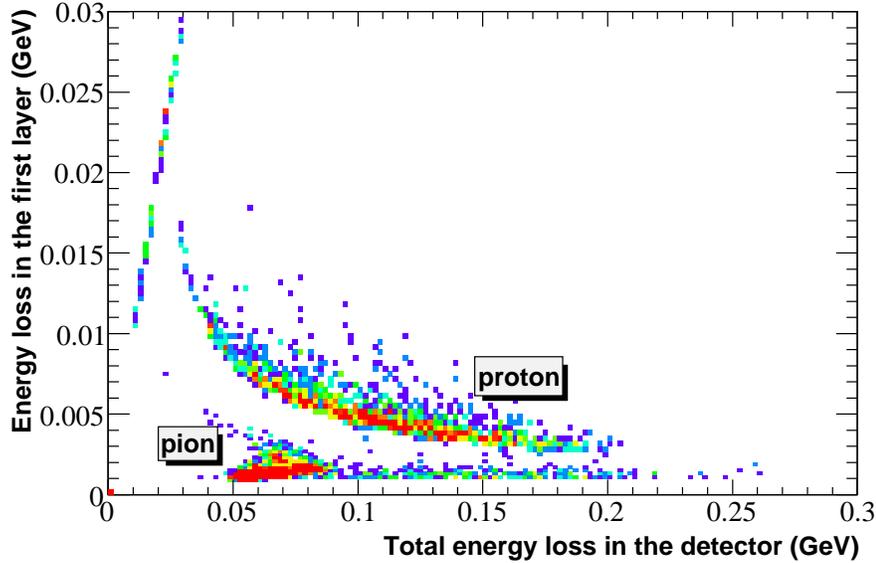}
\caption{A two dimensional plot of $\Delta$E (energy loss in the
inner layer) vs E + $\Delta$E (energy loss in all the layers)
showing the particle separation in ENSTAR. The results are obtained
from GEANT simulations for the events from the $\eta$-mesic nucleus
decay.} \label{nstar}
\end{center}
\end{figure}

    GEANT \cite{GEANT} calculations have been carried out by simulating
the conditions of the real experiment to simulate the ENSTAR
detector's response to eta-nucleus decay particles, namely, protons
and pions. The detector geometry has all its 122 pieces arranged
around the scattering chamber.  The target has been positioned at
the centre of the detector, inside the scattering chamber which is
in vacuum. The existing gap between the various layers of ENSTAR is
filled with air. The $\eta$-mesic nucleus decay events are produced
in a collision of 1.05 GeV proton beam with a target. A Monte Carlo
event generator as detailed in section 2 is used to simulate such
events.  The protons are stopped in the detector while pions, as
expected, pass through it giving only partial energy loss in the
detector. Fig.~\ref{nstar} shows a two dimensional plot of energy
loss in the first layer versus total energy loss in the detector.
The response of various layers of ENSTAR for protons and pions from
such events have been investigated.  The present design does not
plan to obtain full energy information of pions, however, as desired
a mass separation of pions from protons can be achieved. From the
particle selection in the $\Delta$E-E two-dimensional spectrum of
Fig.~\ref{nstar}, the decay events detected within the detector can
be estimated. It is found that the 80 \% of total protons and pions
generated can be identified from the $\Delta$E-E spectrum. It is
further clear from the figure that the energy loss for most of pions
is in the 50-100 MeV range , where a clear separation between
protons and pions can be achieved. The separation of pions from the
protons could be difficult in the higher energy loss region of
pions. However the fraction of the pion events in the energy range
of 100-250 MeV is less compared to number
of events in the low energy range.\\

\subsection{Scintillator grooving and fibre coupling }
\begin{figure}[!ht]
\begin{center}
\includegraphics[scale=0.8]{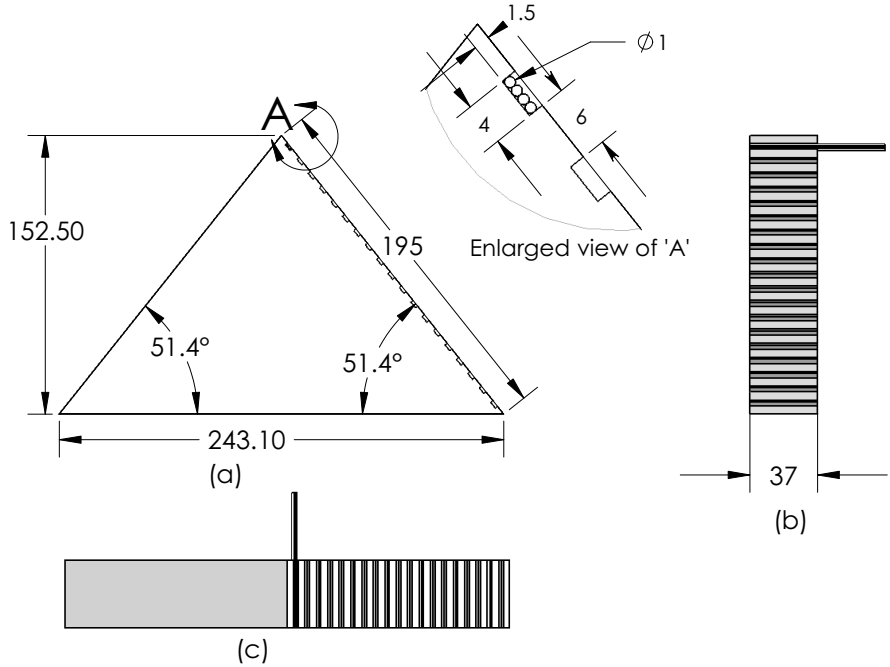}
\caption{The sketch diagram of a typical middle layer scintillator
piece showing the grooves and fibre alignment details. There are 19
grooves on one face of this triangular bar with four fibers placed
inside each groove. The alignment of fibres with the scintillator is
shown in (b) and (c). For illustration purposes fibres in only one
groove are drawn.} \label{sketch}
\end{center}
\end{figure}

         Plastic scintillators, having the properties equivalent to
Bicron BC-408 series, were procured from Scionix Ltd, Netherlands
\cite{scionix}, for the fabrication of detector elements. The  use
of  light  guides   for  scintillator  read out was not practicable
due to the complicated geometry of the detector. The idea of using
wavelength  shifting (WLS)  optical fibres for scintillator read out
was invoked  for  the  present  detector. Earlier
studies\cite{DZERO1,DZERO} have shown that the double-clad fibres
give better light yield (70\% more light) than comparable single
clad ones, due to an increase in the fraction of light that
undergoes total internal reflection. The double-clad WLS optical
fibres having 1mm diameter were used for light transport. A number
of grooves for fixing fibres to the scintillators were made on the
surface of scintillators . The middle and outer layer pieces were
machined for 19 grooves each having 4 mm width and 1.5 mm depth. The
grooves cover roughly 40 \% of the area of one face of scintillator.
For the inner layer pieces, 15 grooves of 1.0 mm width and 1.5 mm
depth were machined with a spacing of 1.5 mm. The machining was done
at  the Central Workshop, BARC using a computer controlled 4 mm (1mm
for the grooves on inner layer pieces) carbide cutter (End-Mill). A
suitable cooling arrangement with chilled air was used in order to
avoid any local heating. Each piece of middle and outer layer has 76
fibres placed in 19 grooves (4 fibres in each groove), while each
inner layer piece has 15 fibres (1 fibre in each groove. The scheme
of fibre scintillator coupling is illustrated in Fig.~\ref{sketch}
for a typical middle layer scintillator piece.\\

\begin{figure}[!ht]
\begin{center}
\includegraphics[scale=1.2]{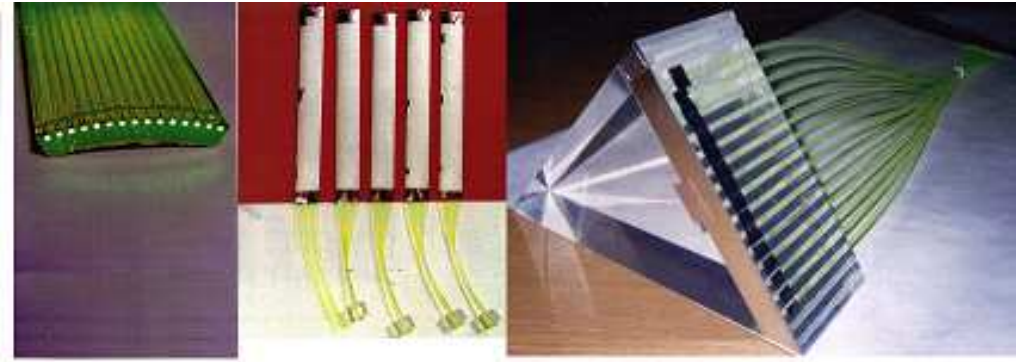}
\caption{Photograph of some inner and  middle layer pieces of ENSTAR
while being fabricated. Left part shows inner layer pieces with
fibres inside scintillator grooves while the middle part shows the
some of the similar pieces after it has been covered with the Tyvek
paper. At the other end  fibres from each piece are bundled together
and coupled to a``cookie''. In the right part of the figure one of
the middle layer pieces with fibres placed inside the grooves is
shown.} \label{scin}
\end{center}
\end{figure}

  The total amount of fibre used was 7.8 km in length.  The fibre
length for each scintillator pieces was decided on the basis of
availability of space in the experimental area. While the length of
fibre  should not be very long in order to minimise attenuation
losses, its bending radius should also be kept high.
 The conventional minimum bending radius  of these fibres is ten times the fibre
diameter. Bending fibres below this radius may result in significant
light loss due to damage in mechanical as well as optical
properties. The length of fibres for each scintillator piece was optimized accordingly.
   Since the light readout is from one end of the fibres only, the light traversing to the
other end must be reflected back.
Therefore, before fixing the fibres, a highly reflective anodized aluminum sheet
(known as EverBrite \cite{EVERBRITE}) was placed on one face of the
scintillator and held in place with aluminized mylar tape.
A good surface finish and polished fibre
ends are essential to prevent light losses at both the reflecting as well as at the readout interface .
This has been achieved by different techniques.
 The cutting and polishing of fibres for the middle and outer layer pieces
were done before fixing them to the scintillators. For polishing,
many fibres were grouped together in bundles inside a perspex tube.
 The fibre face was cut along
 with the perspex by a diamond tipped
cutting tool giving a surface finish of 0.7 $\mu$m. The final
polishing of these fibres was done with 0.3 $\mu$m size alumina
powder on velvet cloth. The polished fibres were fixed in the
scintillator grooves with the Bicron 600 optical cement at few
locations along the grooves. However, to give an additional holding
strength, five-minute epoxy was used wherever necessary. It is
preferable to use the Bicron cement as it has the same refractive
index as that of the scintillator and its light transmission above
400 nm wavelength is more than 98 \%. In addition, aluminized mylar
tape was also used at few places for holding the fibres. For the
inner layer pieces, a different method was followed. First, the
fibres were fixed in the grooves using Bicron cement with a small
amount of five-minute epoxy glue at the ends of the
fibre-scintillator joint. This end of the scintillator along with
the fibres were then polished for all 16 inner-layer pieces. This
was done at the optics workshop of the Spectroscopy Division, BARC
by the lapping technique. Fine alumina powder of 20 $\mu$m, 12
$\mu$m and 6 $\mu$m  were used in successive stages of lapping. The
final finishing was then achieved by polishing with diamond paste
and alumina of 1 $\mu$m and  0.3 $\mu$m sizes giving a surface
finish of 0.3 $\mu$m. Fig.~\ref{scin} (left part) shows the polished
end of one of the scintillator pieces. Finally the highly reflective
EverBrite sheet was placed at this polished end (not shown in the
figure) for light reflection. The other open end of all the fibres
of individual scintillator pieces were bundled together and then
glued to the inside of a 2.54 mm diameter perspex tube - known as
``cookie" \cite{DZERO} (a cylindrical piece of acrylic, matching the
photo multiplier tube in diameter). This end of fibres were polished
along with the cookie. The fibres along with the cookies were
diamond polished by diamond paste and alumina powder.
Fig.~\ref{scin} (middle picture) shows some of the finished (except
for its covering by black foil) inner layer pieces with fibres and
cookie attached. One of the middle layer piece is also shown  in
 Fig.~\ref{scin}(right picture). The cookie end was coupled to the photo-multiplier
tube for conversion of the light signal into photo-electrons which
were then  processed electronically. In order to reduce light losses
from scintillators, the scintillator elements were wrapped with
tyvek, a paper-white reflecting foil made of
polyethylene\cite{TYVEK}. The wrapping by tyvek, apart from light
reflection, also helps in minimising the cross-talk. All the
detector pieces were finally covered by  black tedlar foil for light
tightness and reducing the cross-talk among various detector
elements.\\

\subsection{Scintillator readout details}
 The Bicron optical fibre BCF-91A, used in the present detector
for collecting light produced in the scintillator volume has an
emission spectrum in the visible green region. In order to have an
efficient readout of this light, the photomultiplier tubes (PMTs)
that have a spectral response  extending into the green  region and
which match the light emission characteristic of the wave length
shifting fibres were selected. The PMTs are of the 9112B series
manufactured by Electron Tubes Ltd (ETL), United Kingdom \cite{ETL}.
The PMTs are of 25 mm diameter with Rubidium bialkali photocathode
having an enhanced green sensitivity. A  total of 122
photomultiplier tubes are used for the readout of ENSTAR. The PMTs
have a current amplification  of  10$^6$   and a dark current of
less than 10 nA. The tubes are fast and have a rise time of less
than  3 ns. The PMTs, during the experiment, were covered by
$\mu$-metal sheets which have also been procured from ETL, UK. The
base of the PMTs i.e., voltage dividers are also made by the same
manufacturer. Special aluminum holders were fabricated for holding
the PMTs and cookies together.
\subsection{Detector assembly}
\begin{figure}[!ht]
\begin{center}
\includegraphics[scale=1.]{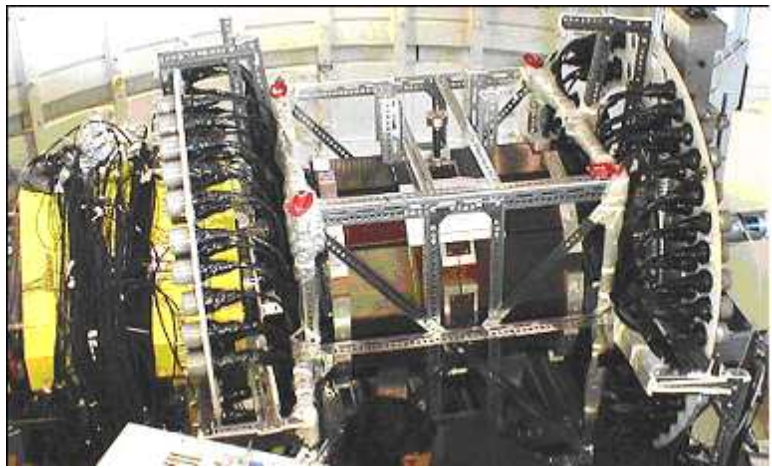}
\caption{A photograph of the ENSTAR detector
 mounted at the COSY beam line along with its support structure and the stand
 which has been used to transport the detector to the beam line. }
\label{full}
\end{center}
\end{figure}

         The pieces of the inner layer of the detector
are the lightest ones and were easy to mount. They were simply held
around the beam pipe/target chamber with tape. The other pieces of
ENSTAR i.e., middle and outer layer pieces are relatively heavier
and special support structures were  designed and built for holding
these pieces in place. The basic supporting structure, which is
mostly an exoskeleton, was made from hylam (low Z material) plates.
Due to the compact geometry of the detector no support structure was
needed inside the sensitive volume of the detector, except only at
few places in the space between the middle and the outer layer where
three support strips made of hylam have been put in each half of the
detector. These support strips were joined by aluminum rings on both
ends for the middle layer. The simulations were repeated with and
without the hylam strips (acting as inactive material inside the
detector). An acceptance loss of less than $\sim$ 1\% for the
particles to be detected is predicted. For the outer layer, the
hylam plates were joined by aluminum brackets at both ends.
 The detector after assembly was placed on a stainless steel stand which was fixed on a
movable trolley made from angle-iron. A stand to hold PMTs was also
constructed and integrated in the same support structure.
Fig.~\ref{full} shows a photograph of ENSTAR detector along with its
support structure mounted at the COSY beam line.\\

\section{Test Measurements}
A number of test measurements were performed during the construction and commissioning
of the detector.
A light-tight black box was constructed for the preliminary tests of the phototubes and the
scintillator pieces. The PMTs and its bases were tested to check for their proper functioning
and to determine their optimum operating voltage.
The variation of signal pulse height from a scintillator tile was studied as
 a function of number of fibres.
The pulse height, which depends on the amount of light collected by
the fibres, is observed to increase with the increase in the number
of fibres and saturates when fibre covers about 30 - 40$\%$ of the
scintillator tile surface. The number of fibres for each
scintillator tile has been optimized accordingly. The light output
of different pieces of ENSTAR was also tested with an alpha source
for which a simple test setup was constructed.\\

\begin{figure}[!ht]
\begin{center}
\includegraphics[scale=0.8]{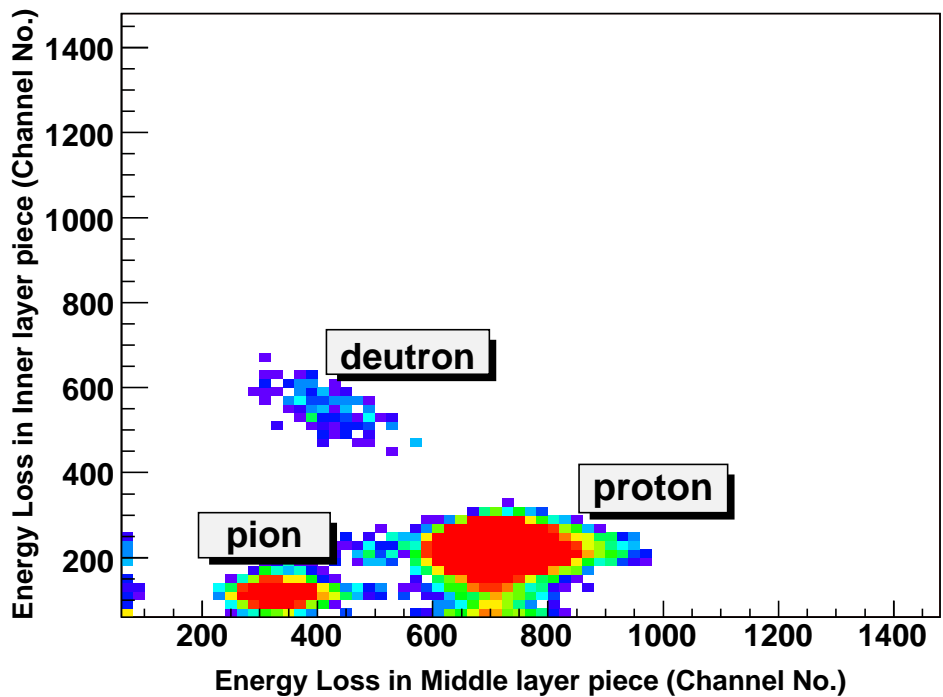}
\caption{ A two dimensional spectrum of the energy losses in inner
and middle layer pieces of ENSTAR mounted at the focal plane exit of
Big Karl magnetic spectrograph in $\Delta$E1 - $\Delta$E2
configuration. The spectrum shows the  pions, protons and deutrons
of energies $\sim$430 MeV, $\sim$ 150 MeV and $\sim$ 80 MeV
respectively, which are selected by a BigKarl momentum setting of
550 MeV/c. }
 \label{ppnew}
\end{center}
\end{figure}

   The first in-beam test at COSY was performed by mounting a few pieces of
ENSTAR from the inner and middle layers arranged in a $\Delta$E1 -
$\Delta$E2 configuration at the exit of the focal plane of the
magnetic spectrograph BigKarl. A proton beam of momentum 1.54 GeV/c,
corresponding to kinetic energy of 865 MeV, was bombarded on a thick
Alumina target. The spectrograph BigKarl was set for different
momenta to select various energies of protons from 35 to 225 MeV and
pions in the range of 150 to 560 MeV. Fig.~\ref{ppnew} shows the
$\Delta$E1-$\Delta$E2 energy spectrum of various particles detected
in scintillator pieces for a typical BigKarl momentum setting of
p/q=550 MeV/c. A Good separation among all particle groups (e.g.
pions, protons, deuterons etc.) was obtained. The particle
identification was confirmed from the time of flight information,
which was measured simultaneously between two hodoscopes layers
placed at a distance 4m apart at the focal plane
of BigKarl.\\

The final test measurement of ENSTAR in fully assembled condition
was performed using a proton beam of momentum 0.870 GeV/c at COSY.
In addition to light-output test of all scintillator pieces, a
study of the relative gain of various elements and absolute
calibration was performed. Several nuclear reactions (pp elastic
scattering, $pp \rightarrow d \pi^+$, proton impinging on a heavy
target etc.) were used for this purpose. Coincidence data i.e. a 2-fold
coincidence between different elements of ENSTAR were collected. In
addition, cosmic-ray data were also
recorded with good statistics.\\

\begin{figure}[!ht]
\begin{center}
\includegraphics[scale=0.8]{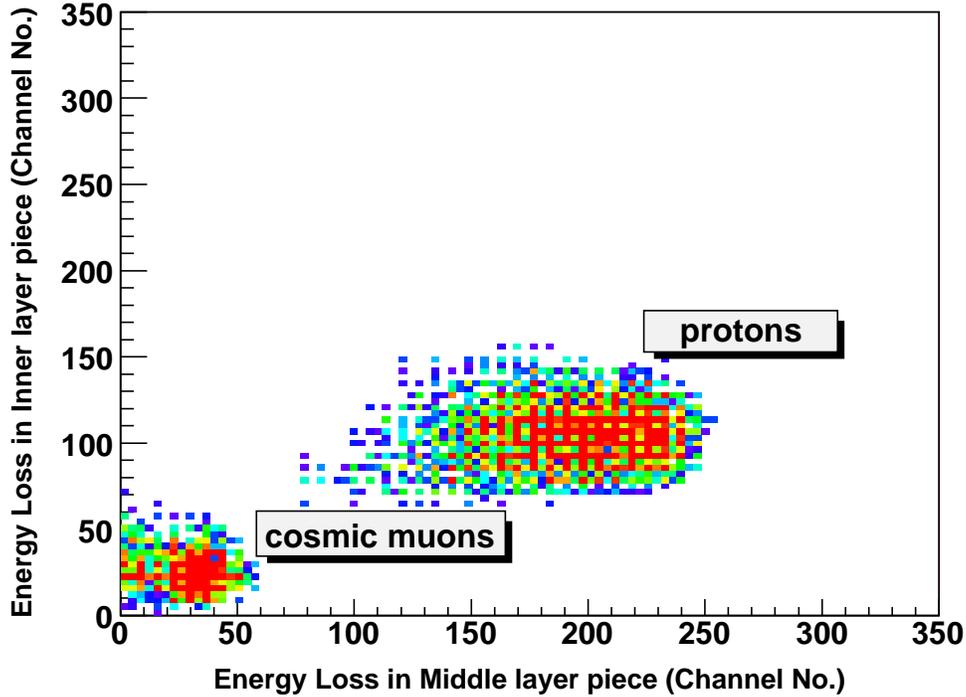}
\caption{ A two dimensional spectrum of one of the inner layer piece
vs the corresponding middle layer piece of ENSTAR for protons in the
energy range of 225-330 MeV obtained from pp elastic scattering
data. Cosmic ray data from a different run are also shown in the
same figure.}
 \label{pp2d}
\end{center}
\end{figure}

In the pp elastic scattering measurement scattered protons having
energies from 25 to 340 MeV are detected in the forward half of the
detector. In this case, the trigger was made from the events which
have a double hit in the inner layer and at least a single hit in
the middle layer. In addition, the condition of co-planarity of the
elastically scattered proton pair was ensured. Light output of one
of the inner-layer scintillator piece versus the corresponding
middle-layer scintillator piece is plotted in Fig. ~\ref{pp2d}. The
proton band shown in the figure corresponds to an energy range of
$225-330$ MeV. A band corresponding to cosmic muons is also shown in
the figure which was obtained from cosmic-ray data collected
separately in a different run,
as described later in the text.\\

\begin{figure}[!ht]
\begin{center}
\includegraphics[scale=0.6]{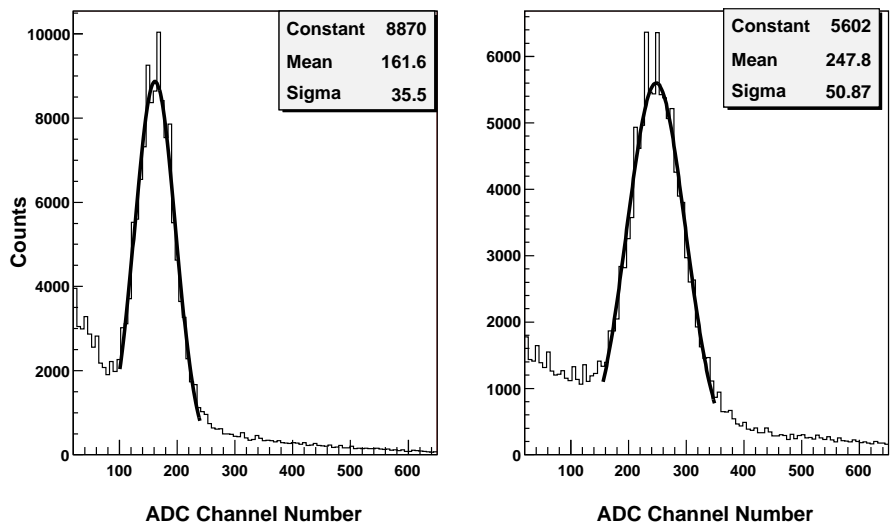}
\caption{ADC spectra of two adjacent middle layer scintillators
uniformly illuminated in $\phi$- direction by scattered particles
from beam impinging on a thick target kept in front of detector.
Peak positions are used to get the relative gains of various pieces
in $\phi$-direction.}
\label{ppheavy}
\end{center}
\end{figure}

\begin{figure}[!ht]
\begin{center}
\includegraphics[scale=0.6]{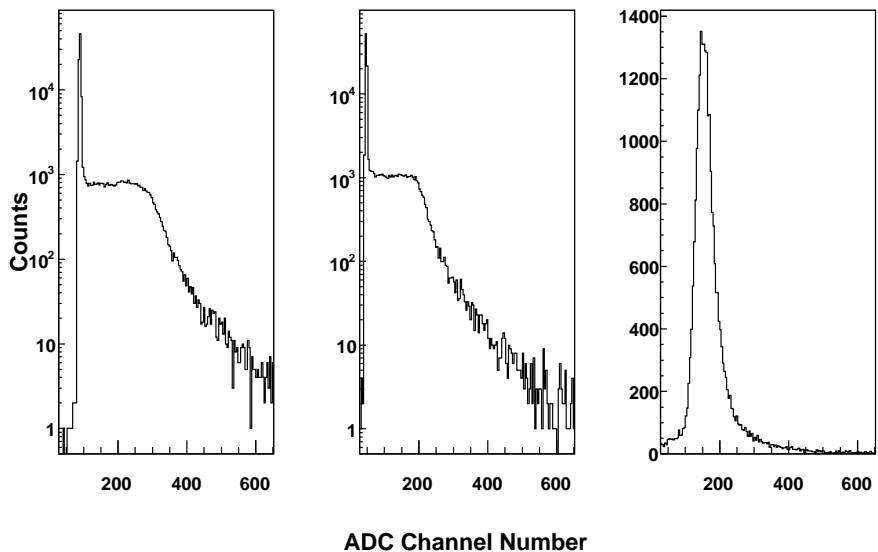}
\caption{ADC spectra from two adjacent middle layer scintillators
for cosmic ray data. The extreme right part of the figure is
generated by demanding an overlapping geometry condition such that
cosmic muons travel a constant thickness. See text for details.}
 \label{cosmicg}
\end{center}
\end{figure}

  The relative gain calibration among different scintillator pieces covering the same $\theta$- but
different $\phi$-ranges is achieved using
reactions in which protons are incident on a thick heavy target. In this case the target,
instead of its conventional location which is at the centre of ENSTAR,
was placed at a position where the beam enters the detector,
Scintillators of the same shape and dimensions form an annular cylindrical ring and therefore,
are uniformly illuminated by the scattered particles.
The relative gain for the  different elements
of the ring is obtained from the peak positions of the spectra shown in Fig. ~\ref{ppheavy}.
   For measurements with the cosmic-rays, two additional
scintillator hodoscope paddles were placed just above and below the
ENSTAR scintillator element being tested. Signals from these paddles formed the cosmic-ray trigger.
ADC spectra from two adjacent scintillators of the middle layer for the cosmic data are shown in
Fig. ~\ref{cosmicg} (left and middle part).  The extreme right part of the figure
is a pedestal subtracted ADC spectrum generated from combination of these two spectra using the
relative gain between the corresponding two pieces as determined above.
 The triangular shape of
middle layer(as well as outer layer) scintillators and the present
detector geometry allow a selection of an overlap portion between
two adjacent scintillators such that muons travel a constant
thickness of 150 mm. The centre of the peak corresponds to an
average energy loss of $\sim$27 MeV since a minimum ionizing
particle typically loses $\sim$1.8 MeV per cm of plastic
scintillator \cite{compact}. This method was used to extract the
absolute
gain calibration of all the middle and outer layer scintillator pieces.  \\

\section{Conclusions}
We have presented a detailed description of a large acceptance
scintillator detector ENSTAR, which  has been designed and
constructed for studying the production and decay of $\eta$-nucleus
bound systems, the $\eta$-mesic nuclei at the multi-GeV hadron
facility COSY. Monte Carlo phase space calculations to simulate the
formation and decay of eta-mesic nucleus predict an energy range of
25 to 250 MeV for the decay protons and energies from 250 to 500 MeV
for the decay pions. The detector is cylindrically shaped in three
layers and is segmented into a number of pieces for the detection of
$\eta$-mesic decay events. GEANT simulations predict a clear mass
separation between the protons and pions based on the energy loss
information in different layers.  A number of test measurements have
been performed to test the performance of the individual components
of the detector. Some of the scintillator pieces have been tested at
COSY by placing them in $\Delta E1-E2$ configuration at the exit of
focal plane detection system  of the magnetic spectrometer BigKarl.
These scintillator pieces have been tested with protons in the
energy range of 35-200 MeV and pions in 150-500 MeV energy range
selected from the Big Karl. The detector has been further tested in
fully assembled condition, using  870 MeV/c proton beam from COSY,
J\"{u}lich. In addition, the measurements using the cosmic muons
have been also performed. For the test measurement with 870 MeV/c
proton beam, the elastically scattered protons having energy in
25-340 MeV range, were detected in ENSTAR. A satisfactorily good
detector response is obtained with the elastic
protons as well as the cosmic muons.\\

\noindent {\bf Acknowledgements}

 This work has been part of the Indo-German bilateral agreement. We are thankful to
the International Studies Division, DAE, India. Our sincere
thanks goes to Dr.S.S.Kapoor and Dr.S.Kailas, BARC for their interest and support
throughout this project.
 We are thankful to Prof. N. K. Mondal, TIFR, Mumbai for many useful discussions and supply of EverBrite sheets.
We are indebted to the technical staff of the Centre for Design and Manufacture, BARC,
Spectroscopy Division, BARC and Institute of Nuclear Physics, Forschungszentrum,
J\"{u}lich for their assistance. We would like to acknowledge the
support from the European community research infrastructure activity under the FP6
``Structuring the European Research Area" programme under the contract no.RII3-CT-2004-506078.

\end{linenumbers}
\end{document}